\documentclass[twocolumn,letterpaper,aps,prc,superscriptaddress,nofootinbib,showpacs,floatfix]{revtex4-1}
\usepackage{graphicx}
\usepackage{comment}
\usepackage{setspace}
\usepackage{amsmath}
\usepackage{amssymb}
\usepackage[textwidth=16cm,textheight=22cm]{geometry}
\usepackage{color}
\usepackage{indentfirst}
\usepackage{xspace}
\usepackage{hyperref}
\usepackage{verbatim}
\usepackage{epstopdf}

\usepackage{lineno}

\newcommand{\ee}{\mbox{$e^{+} e^{-}$}\xspace}
\newcommand{\pp}{\mbox{$pp\,(p\bar{p})$}\xspace}

\begin{document}

\title{Weibull Distribution and the multiplicity moments in \pp collisions}
  \author{Ashutosh Kumar Pandey} \author{Priyanka Sett} \author{Sadhana Dash}

\affiliation{Indian Institute of Technology Bombay, Mumbai, India}

\email{sadhana@iitb.ac.in}

\date{\today}  


\begin{abstract}
A higher moment analysis of multiplicity distribution is performed
using the Weibull description of particle production in \pp collisions
at SPS and LHC energies. The calculated normalized moments and
factorial moments of Weibull distribution are compared to the measured
data. The calculated Weibull moments are found to be in good agreement with the
measured higher moments (up to 5$^{\rm{th}}$ order) reproducing the observed breaking of
KNO scaling in the data. The moments for $pp$ collisions at 
$\sqrt{s}$ = 13 TeV  are also predicted. 
\end{abstract}

\maketitle
\section{Introduction}
\label{intro}

The number of charged particles produced in the final state of a
collision is one of the basic and simplest observable measured in
high energy nucleon-nucleon collisions. Being a global measure to
characterize the final states produced, it provides important insight
and constraints to the mechanism of particle production \cite{KReygers}. 
Therefore, the study of  final state of multi particle production has
generated intense interest in theoretical and experimental high energy
physics. 
Phenomenologically, the multiplicity distributions are expressed in terms of
the probability distribution ($P(n)$) of producing $n$ number of
particles in the final state of a collision. 
If one assumes an independent production of final state particles,
the multiplicity distribution is expected to follow a Poisson
distribution. Any deviation from the Poissonian shape can reveal
more information about the underlying dynamics of particle production
and signal about the existence of multi-particle correlations.
The shape of multiplicity distribution varies with system size and
collision energies and can be incorporated to study the higher moments of the distribution~\cite{prl,sarkisyan,plb}. 

In particular, the study of higher-order moments of the distribution
and its cumulants constitute a sensitive tool to investigate the
existing multi-particle correlations~\cite{kittel,wolf}. 
Although the multiplicity distribution in full phase space is
constrained by global conservation laws, the dynamical fluctuations
arising due to random cascading processes in particle production can
lead to correlations~\cite{ochs,bialas,bialas1,herwig,lund}. 
The deviation from independent production can be
quantified if the factorial moments of order $q$, $F(n)_{q} =
\langle (n (n-1) ..(n-q+1)\rangle / \langle n \rangle ^{q}$  are greater (or less) than unity. These
factorial moments reveal about the dynamical fluctuations. 
The study of reduced moments of order $q$, $C(n)_{q} = \langle n^{q} \rangle/\langle n \rangle ^{q}$  is also
sensitive to see the approximate breaking or holding of KNO
(Koba, Nielsen and Olesen) scaling~\cite{koba}.
The exact KNO scaling would imply the energy independence of the
moments. The violation of scaling was first observed in $p\bar{p}$ collisions
in UA5 experiment~\cite{ua5,ua51}. Recent measurements from CMS and 
ALICE experiment observed strong violation of KNO scaling for higher
order moments ($C_4$ and $C_{5}$) which became
stronger with an increase of energy and width of  $|\eta|$
intervals. The $C_2$ and $C_3$ values were more or less constant
across all energies and  for narrower $|\eta|$ intervals~\cite{cms,alice}.
The energy dependence of higher order moments has been used to improve
or reject different models (Monte-Carlo and statistical) of particle production.
It is well known that multi-particle production processes involve
initial hard scatterings of partons followed by parton fragmentation
and sequential branching. One cannot apply perturbative QCD to the
softer part of particle production and therefore one uses some
phenomenological models or studies based on Monte-Carlo generators.
It is important to see whether the higher order moments and their
variation with beam energy can be understood in terms of some of the 
statistical models or Monte-Carlo generators.  
Previous studies have used the negative binomial distribution (NBD) as 
the most generalized expression for the multiplicity. But deviations
from NBD was observed as the beam energy of the collisions increased~\cite{ua52}.
The choice of Weibull~\cite{weibull,brownpaper} distribution describing the multiplicity
distribution for a broad range of collision energies in both leptonic
and hadronic systems has been quite successful recently ~\cite{weibull1,weibull2}.
Recently, it was shown~\cite{wmoment} that Weibull model was found to reproduce the
genuine correlation measured in $\ee$ collisions for OPAL data~\cite{opal}.   
In this work, the higher order reduced moments and factorial
moments has been calculated using the Weibull description of multiplicity
distribution.  The results have been compared to the multiplicity moments data
measured by different experiments. The idea is to extend the higher
moment calculations to $\pp$ collisions for a broad range of energy and
to see whether one can observe the breaking of KNO scaling using the
Weibull regularity.

\section{Weibull distribution and Moments}
\label{formulation}
The probability density distribution of Weibull for a   continuous random variable $x$ is given by-
\begin{eqnarray}
f(x;\lambda , k) = \frac{k}{\lambda} \left(\frac{x}{\lambda}\right)^{(k-1)}e^{-(\frac{x}{\lambda})^{k}}
\label{e1}
\end{eqnarray}
where $k$ is the shape parameter and $\lambda$ is the scale parameter of the distribution.

The n$^{th}$ raw moment is given by:
\begin{eqnarray}
m_n = \lambda^n \, \Gamma\left(1 + \frac{n}{k}\right)
\label{e2}
\end{eqnarray}
The mean of the distribution, i.e. $m_1$ is denoted by 
$\langle n \rangle$  and is given by, 
\begin{eqnarray}
\langle n \rangle = \lambda \, \Gamma\left(1 + \frac{1}{k}\right)
\label{e2}
\end{eqnarray}
The n$^{th}$ factorial moment of a variable $x$ is defined as:
\begin{eqnarray}
f_n &=& \Big\langle \frac{x!}{(x - n)!} \Big\rangle \nonumber \\
    &=& \langle x(x-1)(x-2)...(x-n+1) \rangle
\label{e3}
\end{eqnarray}
The $n^{th}$ reduced raw moment $C_n$ and factorial moments $F_n$ are  defined as:
\begin{eqnarray}
C_n = m_n/m_1^n ; \,\,\,\,\,\,\, F_n = f_n/m_1^n
\label{e4}
\end{eqnarray}
The first few reduced raw moments for Weibull distribution are given by:
\begin{eqnarray}
C_2 = m_2/m_1^2  = \frac{\Gamma(1 + \frac{2}{k}) }{(\Gamma(1 + \frac{1}{k}))^2}
\label{e6}
\end{eqnarray}
\begin{eqnarray}
C_3 = m_3/m_1^3  = \frac{\Gamma(1 + \frac{3}{k}) }{(\Gamma(1 + \frac{1}{k}))^3}
\label{e7}
\end{eqnarray}
\begin{eqnarray}
C_4 = m_4/m_1^4  = \frac{\Gamma(1 + \frac{4}{k}) }{(\Gamma(1 + \frac{1}{k}))^4}
\label{e8}
\end{eqnarray}
\begin{eqnarray}
C_5 = m_5/m_1^5 = \frac{\Gamma(1 + \frac{5}{k}) }{(\Gamma(1 + \frac{1}{k}))^5}
\label{e9}
\end{eqnarray}

\begin{widetext}
The first few reduced factorial moments for Weibull distribution are given by:
\begin{equation}
F_2 = \frac{\langle x(x-1) \rangle}{m_1^2} 
    = \frac{- \lambda\Gamma(1 + \frac{1}{k}) + \lambda^2\Gamma(1 + \frac{2}{k}) }{(\lambda
\Gamma(1 + \frac{1}{k}))^2}
\label{e10}
\end{equation}
\begin{equation}
F_3 = \frac{\langle x(x-1)(x-2) \rangle}{m_1^3} 
    = \frac{2\lambda\Gamma(1 + \frac{1}{k}) -3 \lambda^2\Gamma(1 + \frac{2}{k})
+ \lambda^3\Gamma(1 + \frac{3}{k})}{(\lambda
\Gamma(1 + \frac{1}{k}))^3}
\label{e11}
\end{equation}

\begin{equation}
F_4 = \frac{\langle x(x-1)(x-2)(x-3) \rangle}{m_1^4}  = \frac{-6\lambda\Gamma(1 + \frac{1}{k}) + 11 \lambda^2\Gamma(1 + \frac{2}{k})
- 6\lambda^3\Gamma(1 + \frac{3}{k}) + \lambda^4\Gamma(1 + \frac{4}{k})}{(\lambda
\Gamma(1 + \frac{1}{k}))^4}
\label{e12}
\end{equation}
\begin{equation}
F_5 = \frac{\langle x(x-1)(x-2)(x-3)(x-4) \rangle}{m_1^5}   \nonumber
\end{equation}
\begin{equation}
= \frac{24\lambda\Gamma(1 + \frac{1}{k}) - 50 \lambda^2\Gamma(1 + \frac{2}{k}) 
+ 35\lambda^3\Gamma(1 + \frac{3}{k}) - 10\lambda^4\Gamma(1 + \frac{4}{k}) + \lambda^5\Gamma(1 + \frac{5}{k})}{(\lambda
\Gamma(1 + \frac{1}{k}))^5}
\label{e13}
\end{equation}
\end{widetext}

\section{Analysis Method and Results}
In this work, we have used the multiplicity distributions of $p\bar{p}$
collisions measured by UA5 experiment~\cite{ua51,ua52}  at SPS energies (200 GeV, 540 GeV and 900 GeV) and $pp$ collisions measured by ALICE~\cite{alice} and CMS~\cite{cms}
experiments at LHC energies (0.9 TeV, 2.36 TeV, 7 TeV and 8 TeV). The analysis method is similar to what has been done previously for higher moment analysis
of \ee collisions at LEP energies~\cite{sarkisyan,wmoment}. The values
of Weibull parameters i.e  $k$ and $\lambda$  were estimated from the measured values of lower order reduced moment, $C_3$ as defined in Eq.~\ref{e7} and the measured mean multiplicity respectively. 
The obtained values of $k$ and $\lambda$ were then
used to calculate the higher order reduced moments as well as the reduced factorial moments. One can also use $C_2$ to extract the $k$ values but the agreement with the higher order moments was found to be better  if $C_3$ was used. 
This is because the higher moments are
very sensitive to $k$. The value of $C_2$ obtained from $k$ parameter estimated from $C_3$ values are in excellent agreement with the measured ones. The analysis was also performed for different pseudo-rapidity ($\eta$) intervals.

\begin{figure*}[ht]
\includegraphics[width=0.9\textwidth]{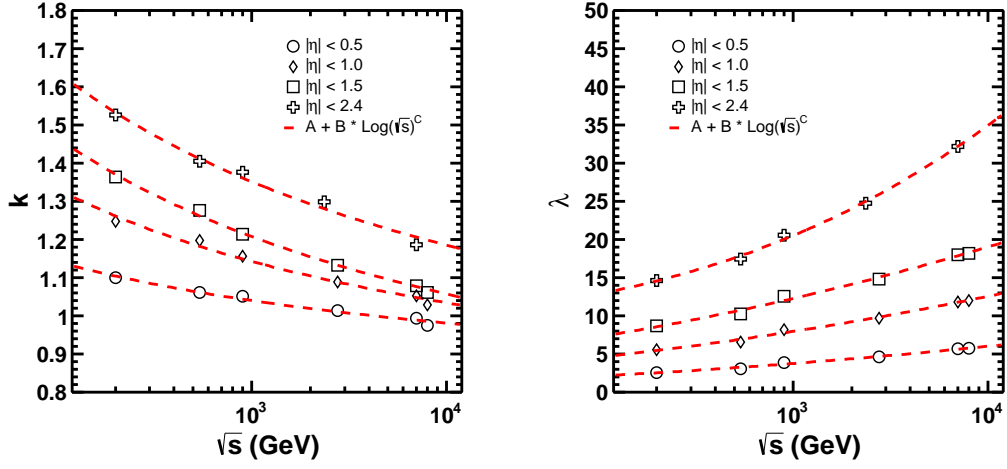}
\caption{The variation of Weibull parameters $k$ and $\lambda$ with center of mass energy in \pp collisions. The variation 
is parameterized by a power law of the form A + B * Log $(\sqrt{s})^C$, shown by the red dashed curve.}
\label{K_lambda}
\end{figure*}

The variation of $k$ and $\lambda$ with collision energy 
and for different $\eta$ intervals are presented in 
Fig.~\ref{K_lambda}. The dependence is described by a power 
law shown by the red dashed curve. 
One can observe that the variation of $k$ and $\lambda$ is consistent with previous multiplicity analysis~\cite{weibull,weibull1}.

\begin{figure*}
\includegraphics[width=2.8in,height=2.8in]{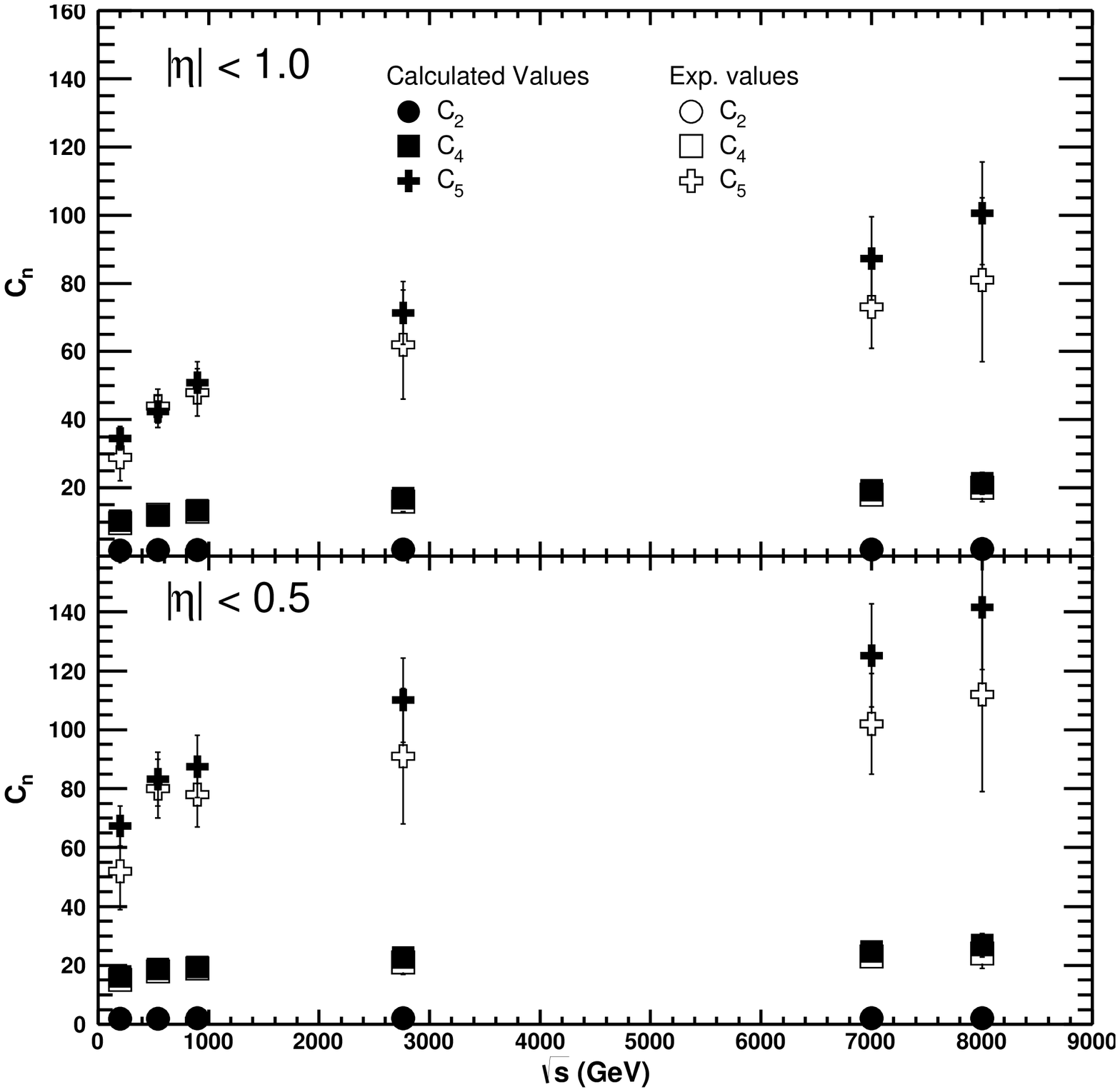} 
\includegraphics[width=2.8in,height=2.8in]{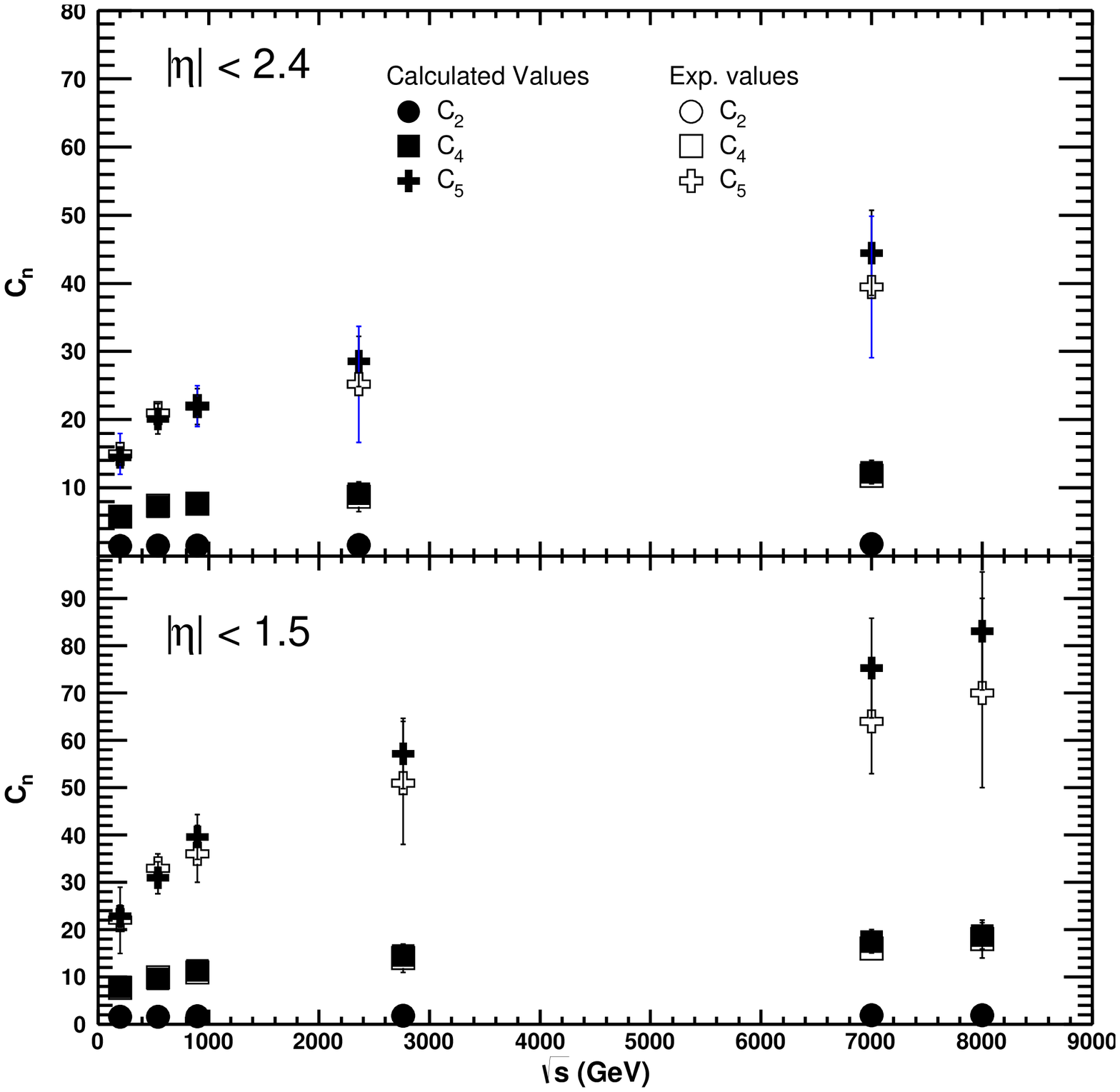}

\caption{The variation of the reduced raw moments (closed 
symbols)	with the collision energies. The Weibull calculations are
compared with experimentally measured normalized raw 
moments (open symbols). 
}
\label{f1}
\end{figure*}

The calculated reduced moments ($C_n$) are shown in different $\eta$ intervals for broad range of energies in Fig.~\ref{f1}. It can be observed from the figure that the obtained Weibull moments agree remarkably well with the measured data within the uncertainties up to
the fifth order for different energies and $\eta$ intervals. 

\begin{figure*}
\includegraphics[width=2.8in,height=2.8in]{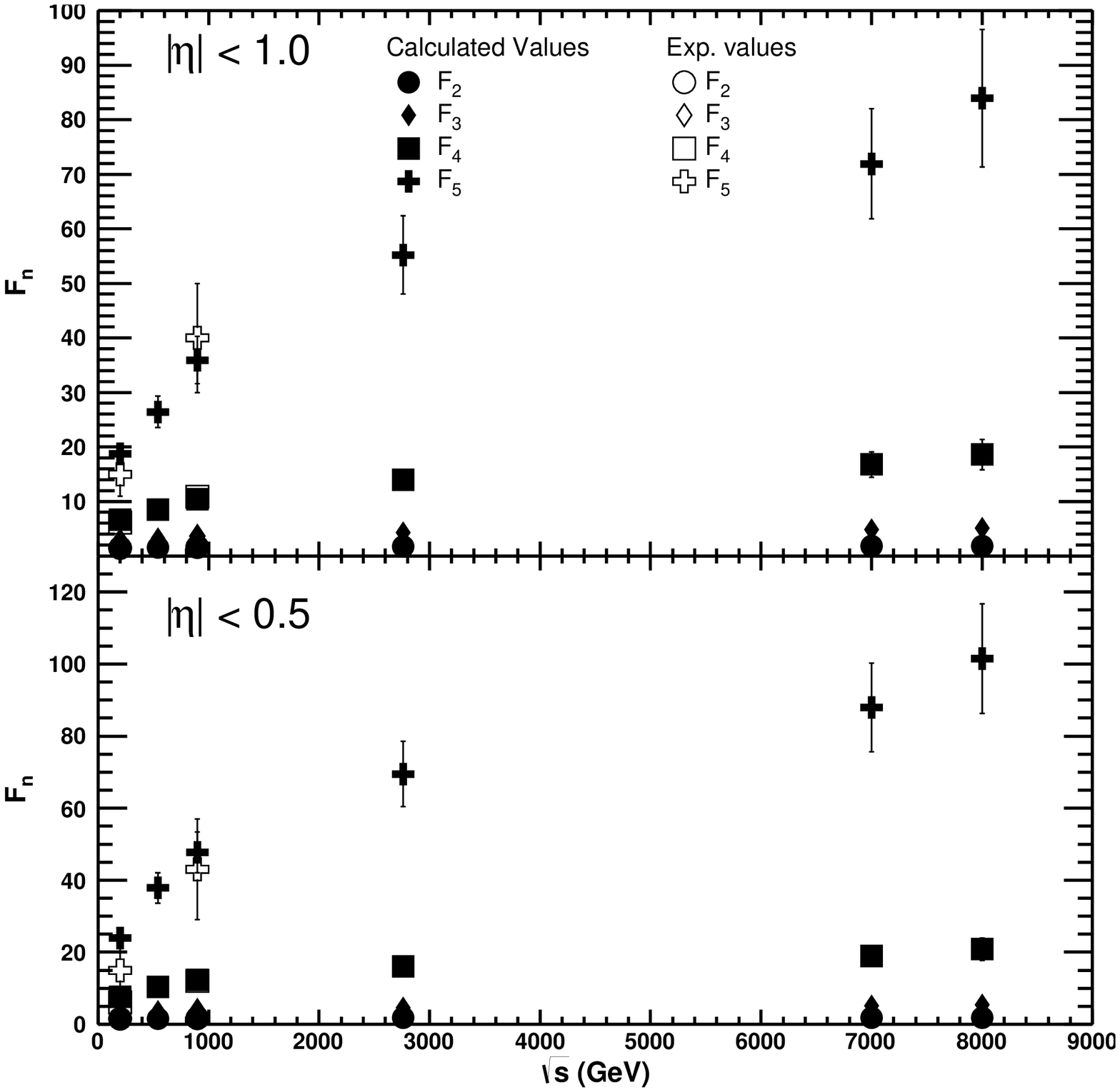} 
\includegraphics[width=2.8in,height=2.8in]{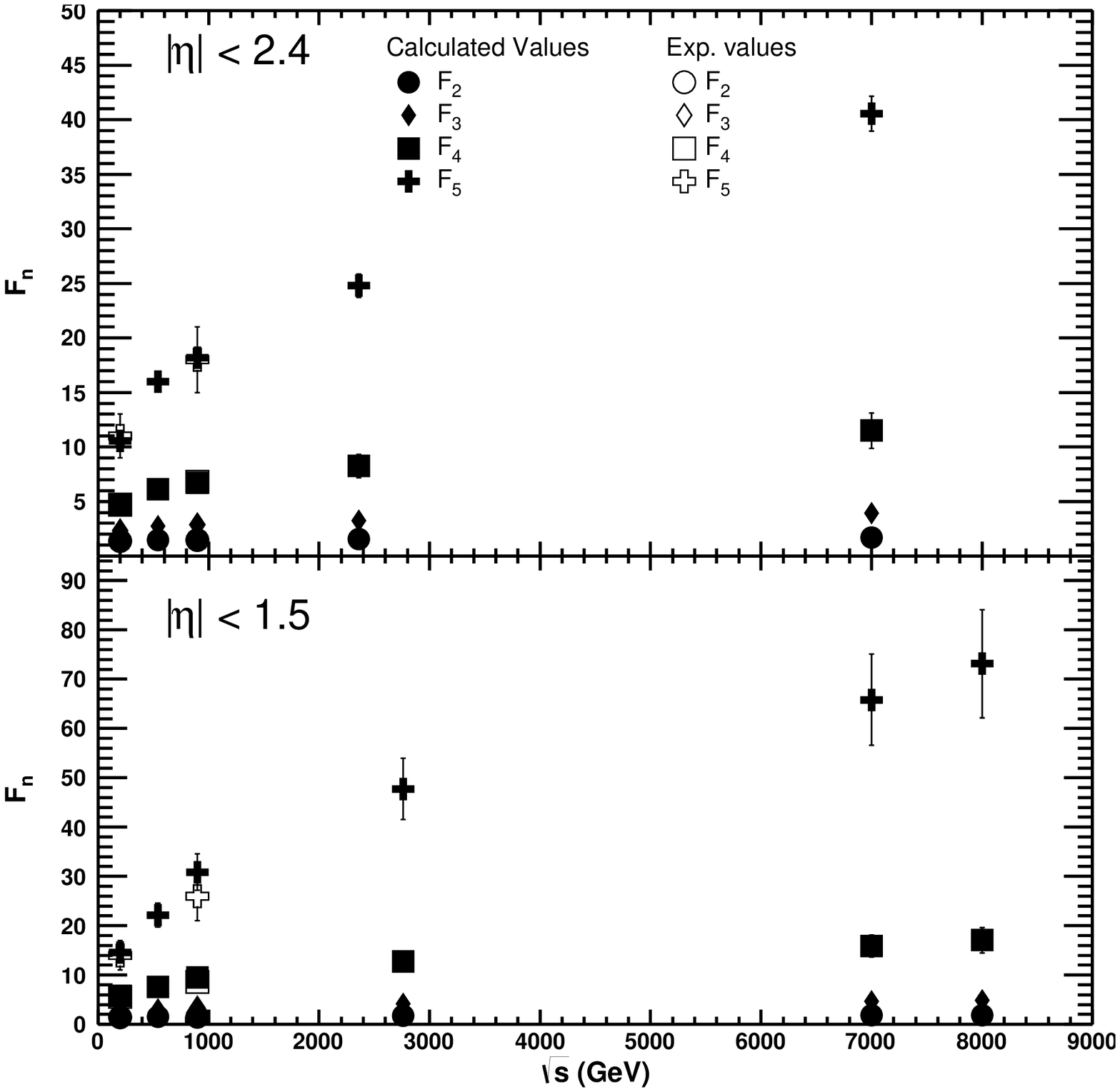}
\caption{The variation of the reduced factorial moments (closed 
symbols)	with the collision energies. The Weibull calculations are
compared with experimentally measured normalized factorial
moments (open symbols) where data is available. 
} 

\label{f2}
\end{figure*}

Fig.~\ref{f2} shows the variation of reduced factorial moments ($F_n$)
with beam energy for different $\eta$ intervals. We need experimental
measurements at LHC energies to see how good the calculations agree. 
The agreement is very nice for measured values of factorial moments by UA5
experiment. The model values confirm the KNO scaling violation which
is seen to increase with beam energy and width of $\eta$ interval.  

The values of $k$ and $\lambda$ for $\sqrt{s}$ = 13 TeV is 
obtained from the parameterization shown in Fig.~\ref{K_lambda}. These  values are used to predict the 
raw and factorial moments at 13 TeV for the above mentioned 
$\eta$ ranges. The values are listed in Table~\ref{t1}. 
\begin{table*}
\caption{The predicted values for mean and other higher moments and factorial moments in $pp$ collisions at $\sqrt{s}$ = 13 TeV.}
\label{t1}
\centering
\begin{tabular}{cccccccccc}
\hline
\hline
$|\eta|$ \,\,\,& \,\,\, $\langle n \rangle$ \,\,\,& \,\,\,$C_2$ \,\,\, & \,\,\, $C_3$ \,\,\, & \,\,\, $C_4$ \,\,\,  & \,\,\, $C_5$ \,\,\, & \,\,\, $F_2$ \,\,\, & \,\,\, $F_{3}$ \,\,\, &  \,\,\, $F_4$ \,\,\,  & \,\,\, $F_5$\,\,\,  \\
\hline	
 0.5 &  6.38 & 2.05  &  6.38  & 26.73  & 140.78 & 1.89 & 5.47 & 21.26 & 103.98\\
  1.0 &  13.03 &  1.95 &   5.65 &  21.62 &  102.68 & 1.87 & 5.21 & 19.13 & 87.22\\
  1.5 &  19.57 & 1.92  &  5.40 &  20.02  & 91.72 & 1.87 & 5.12 & 18.41 & 81.97 \\
  2.4 &  35.15 & 1.73  &  4.19 &  12.90  & 47.81 & 1.71 & 4.05 & 12.20 & 44.26   \\  
\hline
\hline
\end{tabular}
\end{table*}

\section{Conclusion}
The Weibull model of multiplicity distribution was used to calculate
the reduced raw and factorial moments of multiplicity distributions in 
$\pp$ collisions and compared with the experimental data. The obtained
moments (up to 5$^{\rm{th}}$ order) are in good agreement (within experimental
uncertainties) with the multiplicity moments measured for a broad range of energies.
The model predictions have reproduced the violation of KNO scaling as
observed for higher moments in measured data. The predictions for reduced factorial
moments are also made for available $pp$ collisions at LHC energies.
This study further establishes that Weibull distribution seems to be
the optimal statistical model to describe multiplicity distribution as well as the 
higher moments of the same for a broad range of energies in hadron-hadron collisions.

\section{Acknowledgements}
The authors would like to thank Department of Science 
and Technology (DST), India for supporting the present 
work.

\end{document}